# Silicon photonic crystal cavity enhanced second-harmonic generation from monolayer WSe$_2$

**Taylor K. Fryett[1], Kyle L. Seyler[2], Jiajiu Zheng[1], Chang-Hua Liu[2], Xiaodong Xu[2,3,*], Arka Majumdar[1,2*]**

[1] Electrical Engineering, University of Washington, Seattle, WA-98195, USA
[2] Physics, University of Washington, Seattle, WA-98195, USA
[3] Material Science and Engineering, University of Washington, Seattle, WA-98195, USA

* To whom correspondence should be addressed.   Email:arka@uw.edu; Email: xuxd@uw.edu

**Abstract.** Nano-resonator integrated with two-dimensional materials (e.g. transition metal dichalcogenides) have recently emerged as a promising nano-optoelectronic platform. Here we demonstrate resonator-enhanced second-harmonic generation (SHG) in tungsten diselenide using a silicon photonic crystal cavity. By pumping the device with the ultrafast laser pulses near the cavity mode at the telecommunication wavelength, we observe a near visible SHG with a narrow linewidth and near unity linear polarization, originated from the coupling of the pump photon to the cavity mode. The observed SHG is enhanced by factor of ~200 compared to a bare monolayer on silicon. Our results imply the efficacy of cavity integrated monolayer materials for nonlinear optics and the potential of building a silicon-compatible second-order nonlinear integrated photonic platform.

## 1. Introduction

Nonlinear integrated photonics plays a crucial role in building all-optical information processors [1, 2] and novel on-chip light-sources [3]. However, the weak optical nonlinearity of existing material systems results in large optical switching power, rendering optical information processing unattractive. The key to lower the required optical power is to incorporate nonlinear materials onto a nano-scale high quality factor $(Q)$ resonator, where light can be stored in a small volume $(V_m)$ and for an extended period of time [4]. It can be shown that for a nonlinear optical switch, the switching power scales as $V_m/Q^2$ for the third order and $V_m/Q^3$ for the second order nonlinearity [5]. This stronger dependence on cavity $Q$, along with a larger value of second-order $\chi^{(2)}$ nonlinear coefficients compared to $\chi^{(3)}$ coefficients, make $\chi^{(2)}$ nonlinear processes more suitable to realize low-power nonlinear optical devices. Unfortunately, silicon lacks the desired $\chi^{(2)}$ nonlinearity due to its centrosymmetric crystal structure; thus devices based on $\chi^{(3)}$ processes dominate current efforts in nonlinear integrated photonics [3, 6, 7]. While materials with large $\chi^{(2)}$ nonlinearities, such as III-V materials [8] are well-studied, their incompatibility with current CMOS foundries [9] hinders the scalability sought by the integrated photonics community. This is further exacerbated by the fact that deposition of high refractive index III-V materials on silicon changes the optical mode profile significantly, making the phase matching condition more difficult to satisfy. Researchers have also studied aluminum nitride for nonlinear optics [10] and have integrated it on silicon for electro-optic signal processing [11]. However, nonlinear optics with aluminum nitride integrated on a silicon-compatible platform has not yet been reported. A hybrid platform, where we can exploit the scalability provided by silicon photonics as well as realize strong $\chi^{(2)}$ nonlinearity, will be highly attractive for integrated nonlinear nano-photonics with applications to all-optical signal processing.

The recently discovered atomically thin 2D transitional metal dichalcogenides (TMDCs) [12] offer extraordinarily large second-order nonlinear coefficients [13, 14]. They can be easily integrated onto silicon devices by simple van der Waals bonding without the need of lattice matching [15]. In spite of their atomically-thin thickness and evanescent coupling with the light, the effective nonlinearity offered by TMDC-integrated nano-resonator is comparable to that offered by a resonator completely made out of a $\chi^{(2)}$ nonlinear III-V material [16]. Cavity-enhanced second-order nonlinear optics with TMDCs has recently been observed in distributed Bragg reflector cavities [17, 18] and plasmonic resonators [19]. However, both of these cavity systems are unsuitable for low power operation due to their large mode volume and high loss (low Q), respectively. In this paper, we report the enhanced second-harmonic generation (SHG) of a tungsten diselenide (WSe$_2$) monolayer integrated on a planar silicon photonic crystal linear defect cavity, under ~1550 nm laser excitation. The choice of photonic crystal cavity (PCC) is largely motivated by its both small mode volume and high Q[20]. This is particularly important for using exfoliated TMDC monolayers as the modal overlap with the TMDC is further limited by the exfoliated sample size. We measured a cavity enhancement of SHG by a factor of 200, which can be potentially further enhanced by switching to a wide band-gap substrate, such as silicon nitride, and using a resonator with modes at both the fundamental and second-harmonic frequency.

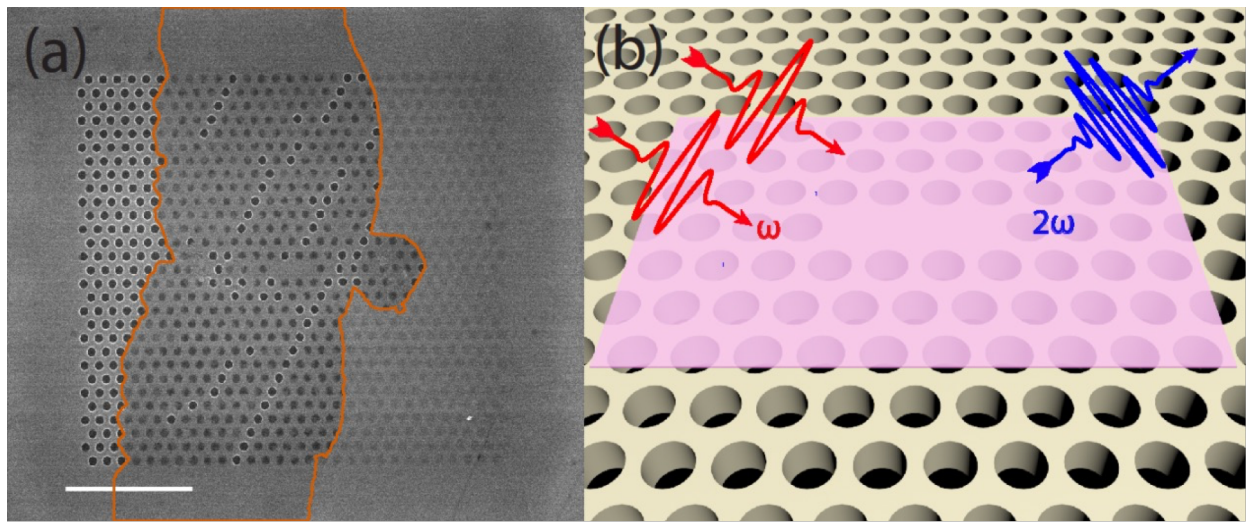

**Figure 1.** The fabricated device: (a) Scanning electron micrograph of the fabricated silicon photonic crystal cavity with monolayer WSe$_2$ on top, indicated by the orange outline. Visible stripes of holes inside the monolayer region are due to ripped monolayer during exfoliation. Scale bar：10 μm. (b) Schematic of the device operation: two infrared (IR) photons denoted by the red Gaussian profiles with frequency ω resonantly couple into the photonic crystal cavity. The cavity mode interacts with the WSe$_2$ (the transparent pink sheet) which then generates the second harmonic photon at a frequency of 2ω, denoted by the blue Gaussian profile.

2. **Methods**

Preparation of the TMDC-cavity device largely followed the standard cavity fabrication and 2D material preparation processes [20-22]. A modified three-hole linear defect (L3) photonic crystal cavity (PCC) [23] is fabricated in a standard 220 nm thick silicon on insulator (SOI) wafer with a lattice periodicity of 398 nm and a radius of 116 nm. We patterned a 250 nm ZEP 520A mask using a 100 kV JEOL JBX6300FS electron beam lithography system. The mask was then transferred onto the silicon by using a chlorine ICP-RIE dry etching recipe followed by an undercutting step using a 1:10 solution of buffered oxide etchant. In parallel to the cavity fabrication, we exfoliated a monolayer $WSe_2$ onto a 300 nm $SiO_2$ on Si wafer. The monolayer was subsequently transferred onto the cavity using a dry transfer method [24]. Figure 1a shows the scanning electron micrograph of an integrated monolayer $WSe_2$ on silicon cavity device.

The cavity modes were identified before and after monolayer transfer using cross-polarized reflectivity measurements (see supplement) [25]. We found several cavity modes in the pristine cavity, including the fundamental mode at 1557 nm with a Q factor of ~10,000 (see supplement). Unfortunately, this high Q mode disappeared after the $WSe_2$ transfer. We have observed severe degradation of high-Q modes in several TMDC-PCC devices. Along with the fundamental mode, we also found several higher-order modes for both the pristine cavity and the 2D integrated cavity (Figure 2a). Exact correspondence between these modes cannot be established with certainty, as the effect of monolayer transfer is not clear. However, the range of quality factor decreased from Q~2000-3000 of the pristine cavity to ~700-800 after $WSe_2$ integration. The degradation of the cavity Q-factor is expected due to the residual loss of 2D material even near 1550nm. We estimate the resulting Q should be in the range of ~1500-1800 (see supplement). The additional Q degradation is attributed to the polymer residues from monolayer transfer [22].

3. **Results**

After measuring the linear spectrum of the $WSe_2$ clad cavity, we moved forward to measure the SHG signal (setup is shown in the supplement). We resonantly pumped the cavity using an optical parametric amplifier (Coherent OPA 9800) to generate light near 1500 nm. The pump laser has a repetition frequency of 250 kHz and pulse width of ~200 fs. We conducted all experiments at normal incidence through a 50X Olympus near-IR objective. The incident light was polarization resolved by passing it through a half- and a quarter-wave Fresnel rhomb and remotely controlled linear polarizer before entering the objective. We detected light near the second-harmonic frequency, where we observed background SHG along with a well-defined cavity peak (Figure 2b). The wavelengths of cavity peaks observed in SHG signal (~745nm and ~758nm) correspond exactly to the half of the cavity wavelength observed in reflectivity (~1490nm and ~1515nm) (Figure 2a). These modes are hereby referred to as mode "α" (at ~1515nm) and mode "β" (at ~1490nm). We verified that the SHG signal appears only when we pump an area with $WSe_2$, and no signal is observed when we pumped the SOI sample without monolayer $WSe_2$. This rules out any possibility of SHG due to the surface nonlinearity [27]. Note that, the wavelength range, where we observe SHG is similar to the range, where $WSe_2$ PL is (Figure 2b). Since PL can be generated from the third harmonic of the laser created via silicon, extra care was taken to ensure we indeed measured the cavity peak in the SHG signal. The PL should have a cubic dependence on the pump power, whereas the measured signal follows a quadratic dependence. Moreover, to observe a cavity signal in PL, we needed to have cavity modes at ~745-758 nm, which was impossible due to the lack of any photonic bandgap in this wavelength range. We also did not observe any such mode under reflectivity, simulation, or PL created by helium-neon laser excitation.

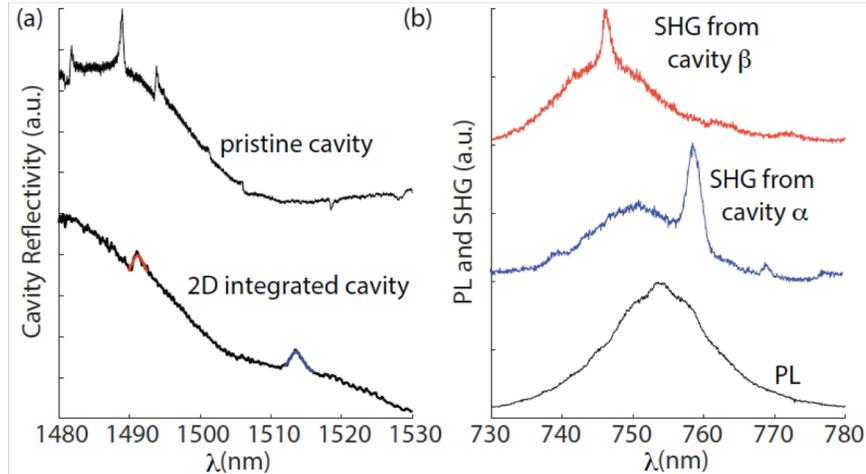

**Figure 2.** Characterization of cavity modes: (a) Cross-polarized reflectivity measurements of the higher order cavity modes before and after monolayer $WSe_2$ transfer. Lorentzian fit is used to estimate the cavity quality factors: blue for the mode "α" at ~1515nm and red for the mode "β" at ~1490nm. (b) Overlay of the photoluminescence signal from $WSe_2$ generated via a HeNe laser (black), SHG spectrum from mode α (blue), and SHG spectrum from mode β (red). We can clearly identify that the cavity-enhanced SHG peaks occur at half the wavelength of the mode seen in reflectivity. The plots are vertically offset for clarity.

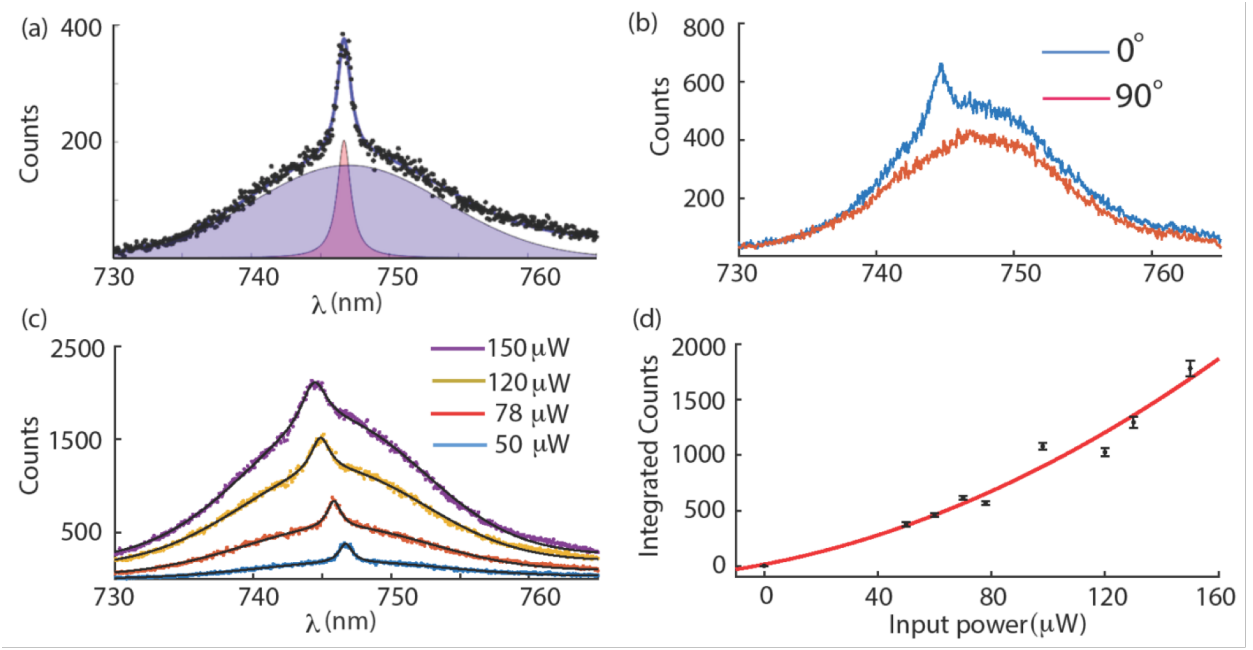

**Figure 3.** Characterization of the second-harmonic signal: (a) A typical SHG spectrum. The total fit is consisted of a Lorentzian for the cavity, a Gaussian curve for the background SHG, and an offset and linear term for the background signal. The corresponding Lorentzian and Gaussian curves are set in to further illustrate their relative contributions to the total fit. (b) Polarization-resolved SHG showing the cavity peak is strongly linearly polarized as expected. (c) Power-dependent (c) SHG spectrum and (d) the cavity-enhanced SHG: each point corresponds to the integrated counts under the Lorentzian fit to the cavity signal. The red line is a quadratic fit.

In the rest of the paper, we analyze the mode "β" while the discussion of mode "α" is in the supplementary materials. To extract the cavity contributions, we fit a Lorentzian to the cavity enhanced SHG peak, a Gaussian to the broad SHG spectrum, and a linear polynomial to the non-SHG background (Figure 3a). The polarization-resolved SHG shows the cavity-enhanced narrow peak is linearly polarized along the cavity mode with near unity degree of polarization (Figure 3b), confirming its coupling to the cavity. We then measured the SHG under different pump powers. Some representative SHG spectra are shown in Figure 3c. All the spectra are also fit with the model described in Figure 3a. A clear blue shift of the cavity resonance was observed, along with a linewidth broadening. We attribute these effects to the free carriers generated by two-photon absorption in silicon [28]. These changes in the cavity parameters are analyzed in detail in the supplementary materials. We plot the area under the Lorentzian fit to the cavity as a function of the input optical power (Figure 3d). A clear quadratic dependence is observed, validating that this signal is due to SHG. At the lowest pump power (~18 $\mu W$), with the least degradation of the cavity due to free carriers, we found a quality factor of ~630. Theoretically, we expect the $Q$ measured in this fashion to be the same as the $Q$ measured from the linear reflectivity spectrum (~745). We attribute the slight deviation from this estimate to the free carrier induced broadening.

Based on our measurements, we estimated the extent of the cavity enhancement by considering the spectral window defined by the cavity full width half maximum. This convention is chosen as the cavity spectral range is much smaller than that of the background SHG signal. We estimated the relative magnitude of SHG from the measured spectrum by integrating over the cavity spectral window for both the Lorentzian (cavity) and Gaussian fits (the background SHG). Hence the nonlinear conversion efficiency for the bare 2D material is given by the ratio of the area under the Gaussian curve and the incident power. For the cavity, the nonlinear conversion is given by the ratio of the area under the Lorentzian curve and the power coupled to the cavity, which is estimated to be ~1% for our experiment (see supplementary materials). The cavity enhancement is thus given by the ratio of the two conversion efficiencies. We found the enhancement to be ~200 at the lowest pump power, which decreases as a function of the pump power (see supplementary materials). We note that previously reported enhancement with DBR cavities are around a factor of 10, primarily due to the lower quality factor (~100) {Day, 2016 #1276}.

## 4. Discussion

Our experiment demonstrated that integrating atomically thin 2D materials onto a photonic crystal cavity results in a two-fold increase in the second-harmonic light in the spectral region of interest. More importantly, we explored a new way to enable second-order non-linear optics in a silicon-compatible platform. We note that the reported enhancement is significantly lower than the theoretical maximum. Partially it is due to that silicon absorbs a significant amount of second-harmonic signal, and the two-photon absorption further degrades the efficiency due to the cavity broadening. In addition, we used a silicon photonic crystal with a mode only at the fundamental frequency. We expect that the SHG can be further enhanced by engineering two cavity modes, one at the fundamental frequency, and the other at the second-harmonic frequency, with good modal overlap between them to ensure phase matching [10, 26]. The theoretical efficiency of the SHG then depends on the quality factors of both cavities $\sim Q_1^2 Q_2$, where $Q_1 (Q_2)$ is the quality factor of the fundamental (second harmonic) cavity mode. Finally, the slab thickness of the cavity and the cavity material itself may play an important role in the second harmonic frequency as recently predicted{Merano, 2016 #1287}.

## 5. Conclusions

Future devices can substantially improve the overall efficiency by using silicon nitride as the underlying material platform. The addition of another cavity mode at the second-harmonic frequency will also provide a considerable performance enhancement. Realizing multiply resonant cavities with good modal overlap in silicon nitride will enable few-photon nonlinear optics under continuous wave operation, as we recently theoretically reported [16]. In addition, second-order nonlinear devices are important for realizing on-chip optical parametric oscillators [3, 29] and optical bistability [1], as well as exploring fundamental studies, including electromagnetically induced transparency [30].

AUTHOR INFORMATION
Author Contributions
A.M. conceived the idea. J.Z. fabricated the SOI cavity. T.F. and C.L. fabricated the 2D material-cavity device. T. F. and K.S. performed the optical characterization. T.F. wrote the paper with input from everyone. X.X. and A.M. supervised the whole project.
Funding Sources
This work is supported by the National Science Foundation under grant NSF-EFRI-1433496, the Air Force Office of Scientific Research-Young Investigator Program under grant FA9550-15-1-0150, and AFOSR (FA9550-14-1-0277). All of the fabrication was performed at the Washington Nanofabrication Facility (WNF), a National Nanotechnology Infrastructure Network (NNIN) site at the University of Washington, which is supported in part by the National Science Foundation (awards 0335765 and 1337840), the Washington Research Foundation, the M. J. Murdock Charitable Trust, GCE Market, Class One Technologies and Google.

ACKNOWLEDGMENT
We thank Mr. Richard Bojko and Dr. Lukas Chrostowski for helpful discussion about silicon photonic fabrication.

**Supporting Information Available**: Further analysis of other cavity modes observed in SHG spectrum, and the effect of free carriers on the cavity are presented in the supplementary materials.

# Supplementary material: Silicon photonic crystal cavity enhanced second harmonic generation from monolayer WSe$_2$

*Taylor K. Fryett[1], Kyle L. Seyler[2], Jiajiu Zheng[1], Chang-Hua Liu[2], Xiaodong Xu[2,3,*], Arka Majumdar[1,2*]*

[1] Electrical Engineering, University of Washington, Seattle, WA-98195, USA

[2] Physics, University of Washington, Seattle, WA-98195, USA

[3] Material Science and Engineering, University of Washington, Seattle, WA-98195, USA

## S1. Characterization of the fundamental mode of the silicon photonic crystal cavity:

We measured the SOI cavity reflectivity by using a cross-polarized reflectivity setup, and found several cavity modes. The fundamental cavity mode has a quality factor of ~10,000. Figure S1 shows the measured spectrum, and Lorentzian fit to the fundamental mode. The spectrum also contains some interference fringes, but we verified by polarization dependence that that peak is indeed coming from a cavity.

*Figure S1: The reflectivity spectrum of the fundamental cavity mode with a quality factor of ~10,000.*

## S2. Theoretical model and simulation of the photonic crystal cavity with 2D material on top:

To understand the effect of the 2D material transfer on the SOI photonic crystal cavity, we performed FDTD simulation using Lumerical FDTD solutions. We use a 0.7 nm 2D material on top of the cavity. We used a non-uniform mesh, with the mesh size being 0.2 nm at the 2D material region, while the mesh becomes 40nm in the photonic crystal. We used dielectric volume averaging for mesh refinement. For the 2D material, we used $n = 4 + 0.1i$ near 1550 nm[1]. The quality factors measured are 150,000 for the pristine cavity, which reduces to 7000. This means the loss due to absorption is $Q_{abs} = 7300$.

**S3. Analysis of the cavity mode a:**

In the main text, we analyzed the performance of one cavity mode, which we referred to as cavity mode "β". We observed similar SHG for other cavity modes as well (cavity mode "α"). Figure S2 shows the pump power dependent SHG spectra of the cavity mode "α". All the diagnostics (power dependence, polarization dependence, Q-determination and resonance frequency) were performed for the mode "α", like mode "β" as explained in the main text.

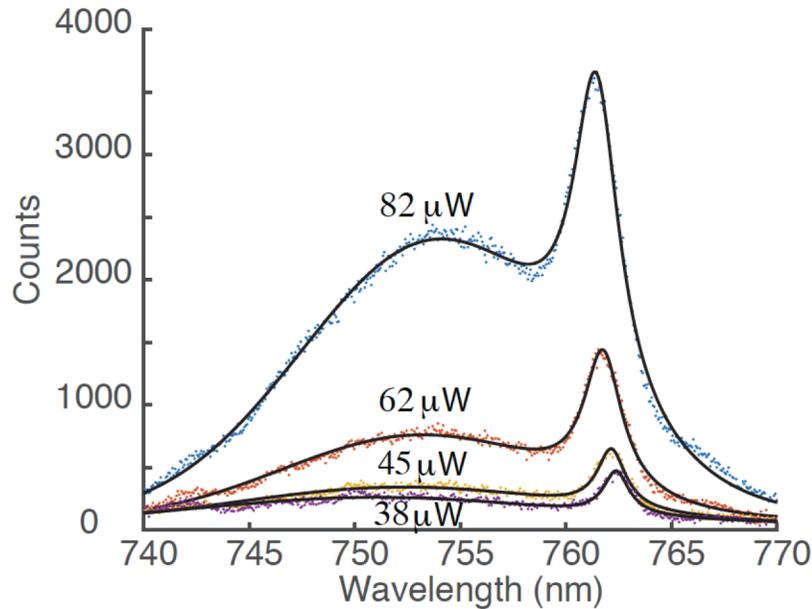

*Figure S2: SHG spectra of the cavity mode "α".*

**S4. Free carrier effects on the cavity:**

As we have shown in the main text, the cavity observed in the SHG spectrum undergoes a blue shift and the linewidth broadens. By fitting the SHG spectrum, we extract the resonance frequency and the cavity linewidth. We attribute this to the free carrier generation in silicon via two photon absorption, which in turn changes the real $(n_1)$ and imaginary $(n_2)$ part of silicon. The change in the refractive index due to free carriers is linear. The resonance frequency and linewidth also vary linearly on $n_1$ and $n_2$. As the carriers are created by two photon absorption, one expects both the resonance frequency and linewidth to be quadratic with the pump power. The linewidth follows such behavior (Figure S3 a). However, the resonance frequency changes almost linearly as a function of the pump power. We attribute the deviation from the expected quadratic behavior to the red shift of the cavity due to thermal heating. This two effects happen simultaneously, and hence we expect a weaker blue shift than the theoretical prediction.

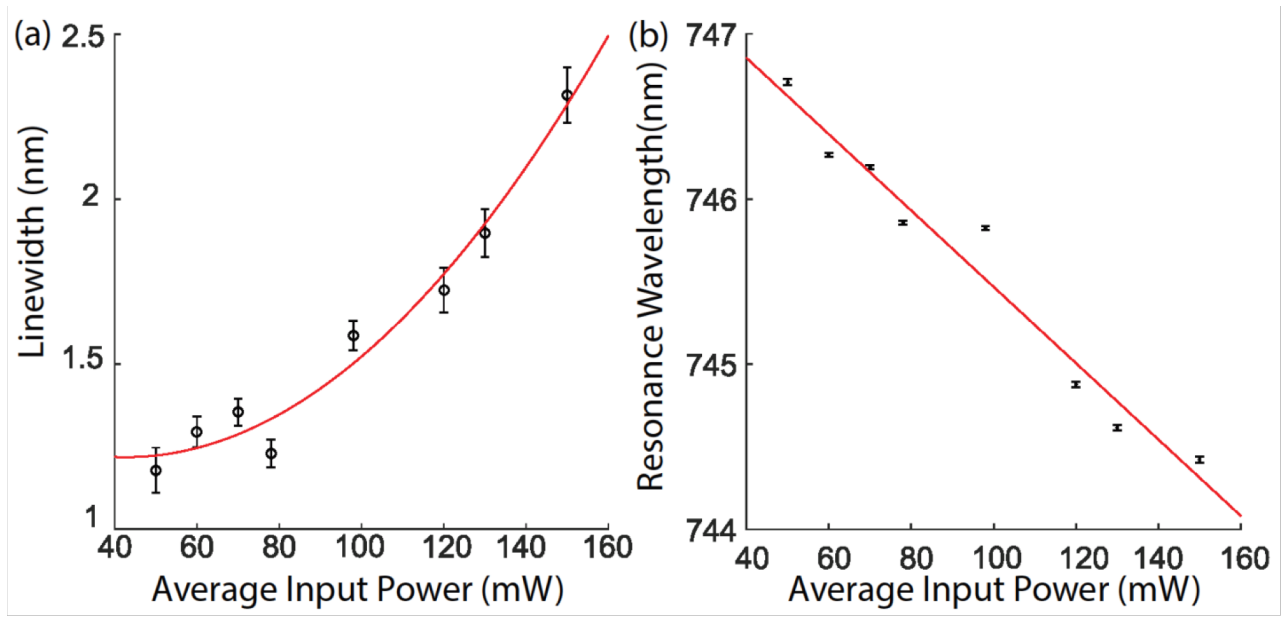

*Figure S3:* *Free carrier effects on mode b: (a) the cavity linewidth and (b) the resonance wavelength as a function of average input pump power. The blue shift of the cavity resonance indicates free carrier effects.*

**S5. Cavity coupling efficiency and the cavity enhancement:**

To estimate the coupling efficiency of the input laser and the cavity, we calculate the overlap integral with the input Gaussian beam and the cavity mode profile. We calculated the cavity mode profile using Lumerical FDTD solutions. Figure S4a inset shows the cavity mode profile. Note that this is a higher order mode. Figure S4a plots the coupling efficiency as a function of the beam radius. We estimate our beam radius to be ~5 µm leading to a coupling efficiency of 1%.

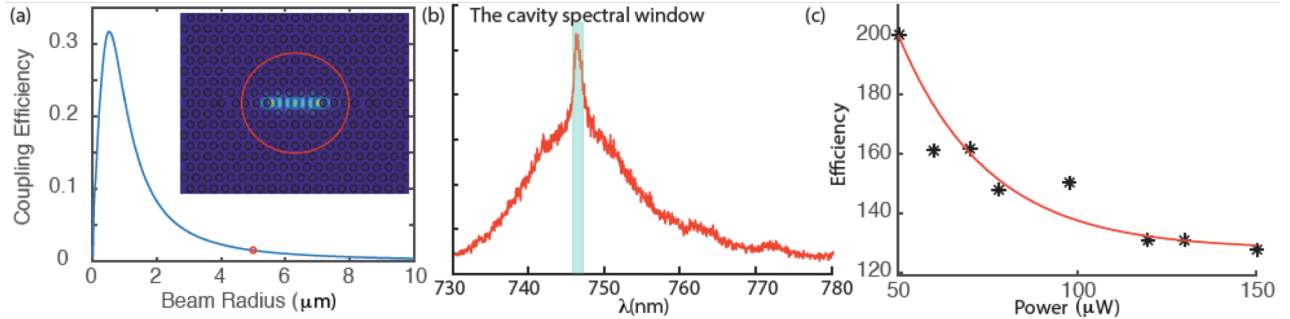

**Figure S4:** *(a) The coupling efficiency as a function of the beam radius. The inset shows the cavity mode profile and the circle indicates the beam-spot size. (b) To calculate the efficiency enhancement of the cavity, we focus on the spectral window where the cavity is resonant. (c) The efficiency as a function of the optical pump power. The red line is a guide to eye.*

Figure S4b shows a typical SHG spectrum, and we highlight the region, over which we integrate to estimate the efficiency and the cavity enhancement. The efficiency depends on the pump power, and Figure S4c shows the efficiency as a function of the pump power.

## S6. Measurement Setups:

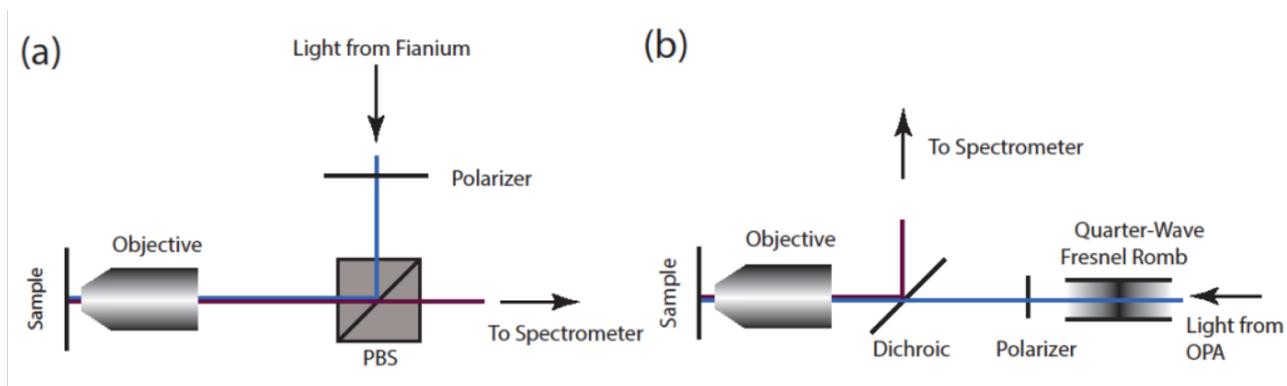

***Figure S5:*** *(a) The cross-polarized setup used to measure the bare cavity quality factors. (b) The polarization dependent SHG measurement setup.*

To measure the quality factor of our nano-resonators we send light from the Fianium supercontinuum is first filtered through a linear polarizer before being incident on a polarizing beam splitter (PBS) to improve the quality of the polarization. The light is then focused down on the cavity which rotates the polarization, and selectively filtered by the PBS before being sent to the spectrometer (Figure S5a). SHG measurements are conducted in a polarization resolved manner, by converting the linear polarized light from the OPA into a quarter-wave Fresnel rhomb to convert it into circularly polarized light, which is then again converted back into linearly polarized light by a remotely controlled linear polarizer (Figure S5b). An IR objective focuses the light down on the sample and then is collected by the spectrometer after being filtered by a dichroic to eliminate the background IR light.